%% file: main.tex
\documentclass[conference]{IEEEtran}
\IEEEoverridecommandlockouts
\usepackage{amsmath,amssymb,amsfonts}
\usepackage{algorithmic}
\usepackage{cite}
\usepackage{courier}
\usepackage{graphicx}
\usepackage{stfloats}
\usepackage{tabularx}
\usepackage{textcomp}
\usepackage{xcolor}
\usepackage{subfiles}
\def\BibTeX{{\rm B\kern-.05em{\sc i\kern-.025em b}\kern-.08em
    T\kern-.1667em\lower.7ex\hbox{E}\kern-.125emX}}
\begin{document}

\title{An Overview on Cloud Distributed Databases for Business Environments}

\author{\IEEEauthorblockN{1\textsuperscript{st} Allan Vikiru}
\IEEEauthorblockA{\textit{School of Computing and Engineering Sciences} \\
\textit{Strathmore University}\\
Nairobi, Kenya \\
allan.vikiru@strathmore.edu}
\and
\IEEEauthorblockN{2\textsuperscript{nd} Mfadhili Muiruri}
\IEEEauthorblockA{\textit{School of Computing and Engineering Sciences} \\
\textit{Strathmore University}\\
Nairobi, Kenya  \\
mfadhili.muiruri@strathmore.edu}
\and
\IEEEauthorblockN{3\textsuperscript{rd} Ismail Ateya}
\IEEEauthorblockA{\textit{School of Computing and Engineering Sciences} \\
\textit{Strathmore University}\\
Nairobi, Kenya \\
iateya@strathmore.edu}
}

\maketitle

\subfile{sections/abstract}

\subfile{sections/introduction}

\subfile{sections/dist-db}

\subfile{sections/cloud-comp}

\subfile{sections/cloud-db}

\subfile{sections/conclusion}

\subfile{references}
\end{document}

%% file: sections/abstract.tex
\begin{abstract}
Cloud-based distributed databases are a popular choice for many current applications, especially those that run over the Internet. By incorporating distributed database systems within cloud environments, it has enabled businesses to scale operations to a global level, all while achieving desired standards of system reliability, availability, and responsiveness.


Cloud providers offer infrastructure and management tools for distributed databases as Database-as-a-Service (DBaaS), re-purposing the investment by businesses towards database services. This paper reviews the functionality of these services, by highlighting Amazon Relational Data Service (RDS), suited for handling relational distributed databases.

\end{abstract}


%% file: sections/introduction.tex
\section{Introduction}
\label{intro}

Numerous systems run on distributed database systems hosted on cloud architectures. Facebook applies several databases such as MySQL, Apache Hadoop and Apache Cassandra which are distributed across data centres. \cite{shivang} Due to the amount of data generated by users streaming video, Hulu applied Apache Cassandra to ensure their content is infinitely scalable and always available. \cite{verge} The Weather Company applies three database products: Riak to handle scaling, Cassandra to distribute data to the company and third-party weather applications, and MongoDB to distribute to their website and mobile applications, all hosted on Amazon Web Services. \cite{henschen}

Distributed cloud database systems are multiple data storage sites hosted in different geographical locations and interconnected over the Internet. Data is readily available to users due to this dispersion, and their access and management are carried out by virtual services offered by cloud providers. \cite{mathur} Different types of cloud-based distributed databases exist with various data models \cite{vlasceanu,kumar}:
\begin{itemize}
    \item Relational databases where data is modelled as tables with rows and columns, and data is queried using Structured Query Language (SQL) language.
    \item Document databases - data is stored in JavaScript Object Notation (JSON) format and queried by keys or filters, similar to that used in applications.
    \item Key/value databases involve data being stored as attributes based on a unique identifier, which is used for querying.
    \item Graph databases include data being stored as nodes that are interconnected by edges. Querying is done by a graph query language.
    \item Time-series databases where data is ordered by timestamp and queried by SQL or any other query language.
\end{itemize}

There has been a considerable shift by organisations to applying distributed designs in their data storage, compared to centralized architectures. Reference \cite{coronel} attributes the  reconsideration for organisational database requirements to the need for accessing data from dispersed business units for rapid decision making. Small-to-medium sized healthcare institutions lack the adequate resources to effectively manage centralised systems, prompting them to set up distributed data services on the cloud which improves the scalability and availability of data. \cite{pedrosa} It is emphasised that businesses operate at the scale of distributed storage on cloud environments to meet the ever-increasing number and expectations of customers, such as remote accessibility, instant responsiveness, and continuous engagement. \cite{marston} Moreover, it is predicted that single instance databases will fail to meet requirements for normal operations; especially in e-commerce where factors such as load time for pages and images play a huge role in determining customer engagement. \cite{ploetz}

This paper thus, looks into the application of distributed data storage within cloud environments, with an insight on how they can be implemented within standard business environments. The rest of the paper is divided into three major sections; the first documents the functionality of distributed database systems, concepts behind their implementation and design, and the process of transaction management. The next section looks into cloud computing and its respective application in business together with models for service delivery, including distributed cloud architectures. Lastly, an insight into AWS Relational Data Service (RDS) is provided, a relational cloud distributed data storage service offered by Amazon Web Services. This includes a description of the service, a comparison to similar cloud-distributed database services in the market, and a sample application on how a database is controlled using AWS RDS.

%% file: sections/dist-db.tex
\section{Distributed Databases}
\label{distributed-db}

\subsection{Definition}
A distributed database system can be defined as an integration of autonomous local databases that are geographically distributed and interconnected by a network, effectively controlled by a distributed database management system (DDBMS). \cite{cellary}

\begin{figure}
\centering
\includegraphics[scale=0.45]{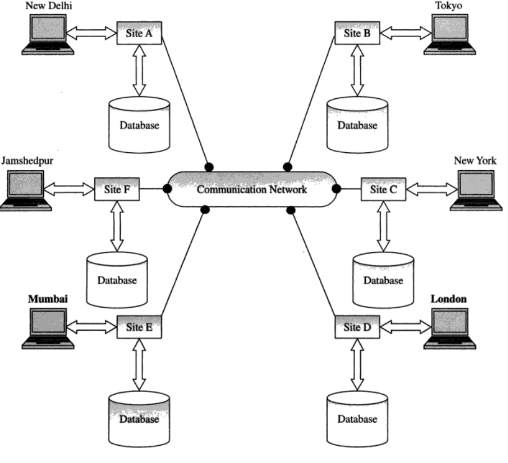}
\caption{Distributed Database System Architecture \cite{ray}}
\label{fig:dist-db-systems}
\end{figure}

 The functionality of distributed database systems is illustrated in Figure \ref{fig:dist-db-systems}. Nodes from different physical locations are connected to their respective sites for processing queries related to both local and remote databases. The sites, in turn, are connected to their corresponding local databases and the underlying communication network that connects all system nodes. Connected nodes are heterogeneous in terms of hardware, software, network configurations and data, and do not share any computing resources among each other such as memory and disk storage. It is paramount that information in all databases within the distributed system network be logically related. \cite{elmasri, singh} 

\subsection{Concepts of Distributed Database Systems}
\subsubsection{Transparency}
Reference \cite{coronel} defines transparency as the view of a distributed database as a centralised logical view to the system user. By establishing full transparency, the user is unable to carry out data transactions without handling the distribution of data, its partitioning, and location and replication of the database partitions. \cite{naik-15} Transparency can also be achieved in the following forms \cite{kumar-13, naik-18}:

\begin{itemize}
    \item Access transparency - Users cannot differentiate access methods to either local or remote databases. 
    \item Migration transparency - Users are unaware of movement of any system information or processes as their operations are not interfered with.
    \item Scaling transparency - Additional system resources or nodes can be appended to or excluded from the system without affecting normal operations
    \item Performance transparency - Reconfiguration of the system can take place without interfering with user operations.
    \item Database transparency - Users can access data from the system databases without knowledge of commands.
\end{itemize}

\subsubsection{Autonomy}
Autonomy refers to the degree at which individual database management systems or nodes can operate independently. This allows for local transactions and configurations to be carried out without affecting operations of the larger distributed system. Three factors that autonomy can apply to a system node are in its design i.e., freedom to apply the data model and transaction management techniques of preference; the communication with other nodes, where it decides what information to share with another node; and lastly its execution, which encapsulates its ability to carry out transactions in its most preferred manner. \cite{elmasri, oszu}

\subsubsection{Reliability \& Availability}
Reliability is defined as the ability of a database system to perform consistently without errors, while maintaining the safety, recoverability, and integrity of data whereas availability is the probability of a system to be continuously accessible by users during a time interval. \cite{elmasri,oszu} 

Regarding distributed database systems, availability is covered by creating and maintaining copies made of data items across nodes. The number, locations, access, and consistency of replicas are metrics that are used to determine the measure of availability when designing a database system. Conversely, database reliability is based on verifying the expected and actual results produced after executing transactions, handling concurrent write accesses to data items and maintaining consistency among clients when accessing the same data item. \cite{domaschka}

\subsection{Distributed Database Design}
\subsubsection{Fragmentation}
This technique involves dividing the database into several subsets and storing the partitions in various sites. In this, a relational database table can be fragmented into partitions for purposes of physical storage and developing copies. The partitions are independent and allow for reconstruction into the original table. Application of fragmentation to a database creates the concept of fragmentation transparency, implying users are presented with a view of the data where the fragments are recombined logically. The system optimiser is a module for determining how the partitions are accessed to carry out queries. \cite{naik-15, singh}
There are three strategies for database fragmentation \cite{coronel, truong}:

\begin{itemize}
    \item Horizontal fragmentation, which divides a table into subsets based on the rows. It is mostly done when a single machine cannot handle the amount of data or loads presented when querying the data.
    \item Vertical fragmentation, which divides a table into subsets based on the columns. This is determined by the properties of the data store and the intended goal for usage optimisation.
    \item Mixed fragmentation, where both horizontal and vertical fragmentation are combined. A relational table may be divided into several horizontal partitions, that each have subsets of table columns.
\end{itemize}

Reference \cite{truong} presents some issues that may arise during partitioning, such as fragments being unequally sized, some partitions being more frequently queried than others, and complicating the query processing, affecting their speed and inconsistency. These can be effectively handled by tailoring the partitioning strategy as well as the schema design when converting from centralised to distributed data stores.

\subsubsection{Allocation}
Data allocation is the process of deciding the location of sites for data hosting. There are three strategies for this \cite{coronel, singh}: 
\begin{itemize}
    \item Centralised strategy, where the entire database is stored in a single site, despite users being geographically distributed. Communication costs are high since all other sites except the central site need the network to refer to the database. Also, in case of failure, the entire database system is lost since it is entirely hosted in a single site, affecting the reliability and availability.  
    \item Partitioned strategy, where the database is fragmented and stored at different locations. In this there are low costs for storage, since there are no copies created, and communication costs are lowered as query loads are distributed among the sites. Reliability and availability are still low as a crash occurring at one site still leads to data loss, however it is higher compared to the centralised strategy.
    \item Replication strategy has copies of one or more database fragments stored at various sites. This maximises the availability and reliability but increases the costs for storage and communication. Algorithms for allocation consider a variety of factors in determining the location such as goals for performance and availability, database size, types of transactions and disconnected operations for mobile users.
\end{itemize}

\subsubsection{Replication}
Data replication allows for storage of data in more than one physical location, increasing the availability of the system. This concept supports replication transparency, where users can interact with the system without necessarily being knowledgeable about the copies of the database. \cite{oszu} Replication transparency simplifies the process of creating and destroying database copies to meet system requirements without affecting user activities. The system optimiser module is also responsible for determining the replicas to be accessed in the process of query executions. \cite{singh}

Reference \cite{elmasri} cover three cases of replication: full replication where the entire database is copied across all sites, enhancing the performance especially for global transactions since it eliminates the need for cross-site communication. Conversely, it reduces efficiency of update queries as a single logical update must occur at all sites. No replication is the second case, where each fragment is stored in one site. This maximises costs for cross-site communication but reduces those for concurrency control and recovery as the efforts are only limited to the single copies per site. Lastly, partial replication involves developing copies of some database fragments. This is heavily applied in instances of field workers in an organisation who carry on their tasks with a replica and later synchronise them to the primary database. 

Application developers need not worry about the consequences of data replication but more on the location of data and the time for replication. \cite{truong} The location is crucial to minimise network latency of updates that can be affected by site failures and network partitions. The time for replication can be classified as synchronous, where data is copied to all replicas before responding to the request sent by users, or asynchronous where data is stored on one replica before responding to a request.

\subsection{Transaction Management in Distributed Database Systems}
Transactions in a distributed database system are sequences of Read and Write operations that take the database from one reliable state to another, ending with either of two statements: a Commit, indicating the verification of all operations performed by the transaction; or a Rollback or Abort, which indicates the cancellation of operations. There are two types of transactions: query transactions which consist of only Read operations that do not modify data objects but only access them and returns values to the user. Whereas, update transactions consist of both Read and Write operations, allowing for both access and modification of data objects. \cite{ahmad,ezechiel}
Reference \cite{tok} states that managing distributed transactions involves ensuring database consistency and reliability when handling both local and global transactions, and in the case of any communication or on-site failures. To manage consistency and reliability, it is essential that transactions maintain these four properties \cite{ahmad,ezechiel}:

\begin{itemize}
    \item Atomicity: This dictates that all actions related to a transaction are completed or none is carried out at all, therefore treated as a unit of operation. For instance, in the case of a crash, the system should complete the remainder of a transaction, or all transactions will be undone. 
    \item Consistency: This deals with maintaining concurrency control, which avoids data from being modified by a transaction that has not been committed.
    \item Isolation: It ensures that execution of a transaction occurs completely independent of other transactions i.e., no other query or update transactions should execute while another is occurring, enabling each transaction to manipulate a consistent database.
    \item Durability: It guarantees that the updates made by a transaction stay permanent. In case the system crashes or aborts a transaction, results once committed are not modified or undone.
\end{itemize}

\begin{figure}
\centering
\includegraphics[scale=0.45]{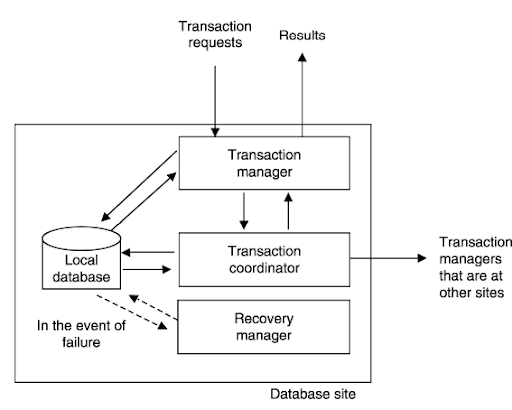}
\caption{Transaction Management Model \cite{tok}}
\label{fig:transaction-mgt}
\end{figure}

In a distributed database management system, there are four interconnected modules that work together to ensure effective transaction management. First, the transaction manager at each individual database manages execution of both local and global transactions initiated at a particular site. It implements a concurrency control mechanism to coordinate concurrent execution of transactions. The transaction coordinator at each site plans and schedules sub transactions that are executed on multiple sites. Plus, it determines the results of sub transactions i.e., if they are committed or aborted. Third, the recovery manager is responsible for maintaining database consistency in case a failure occurs, while the last module, a buffer manager, handles the efficient transfer of data between disk storage and main memory. Figure \ref{fig:transaction-mgt} describes the interaction between these modules at each database site. \cite{ray, tok}

Reference \cite{ray} describes several issues that transaction management subsystems should consider to meet its function of enhancing database consistency:
\begin{itemize}
    \item Improvement of CPU and main memory utilisation for database applications. This focuses on usage of minimum resources for input/output operations.
    \item Minimal response time for individual transactions, especially in the case of global transactions where cross-site communication occurs.
    \item Maximum availability for transaction recovery and concurrency control. 
    \item Minimum communication cost specifically in the execution for global transactions. It is crucial for the transaction manager to adopt preventive measures when transferring data and messages for controlling execution of a global application.
\end{itemize}

%% file: sections/cloud-comp.tex
\section{Cloud Computing Architecture}
\label{cloud-comp}

\subsection{Definition}
Cloud computing is the utilisation of web-based devices, resources such as networks, servers, storage, and services by system developers to implement systems that run over the Internet and are hosted in large-scale data centres. The resources are virtual, implying that they can be dynamically provided, reconfigured, and used based on a pay-per-use economic model. Moreover, the service provider need not provide much effort in leasing and managing resources to consumers; allowing them to focus more on managing the hardware hosting the services. \cite{hajibaba, jamsa}

References \cite{hajibaba} and \cite{hayes} describe an abstract view of the cloud computing environment, where high-performance machines hosted in data centres are connected over high bandwidth networks. Once end users are connected to the Internet, they can interact with the system by sending requests, which are then coordinated by the Cloud manager that redirects to the appropriate site for processing. Upon processing, the results are relayed back to the Cloud manager which eventually presents them to the user.

\begin{figure}
\centering
\includegraphics[width=0.8\columnwidth]{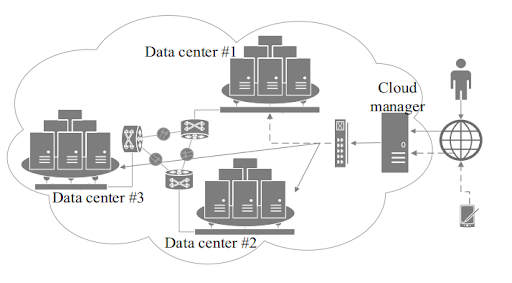}
\caption{Cloud Computing Architecture \cite{hajibaba}}
\label{fig:cc-architecture}
\end{figure}

The shift to cloud computing by organisations is driven by several factors \cite{marston, ting, lisdorf}:
\begin{itemize}
    \item Economics is a major contributor, with many organisations working towards cutting down expenses for infrastructure and system maintenance.
    \item Data and system security also influence the adoption of cloud computing technologies, as services continually upgrade their systems against threats such as denial of service attacks. Moreover, some security features such as encryption and identity and access management are established by default, which would be difficult to implement from scratch.
    \item Trends around remote working and Bring-Your-Own-Device (BYOD) to enhance worker flexibility is greatly facilitated by cloud computing services which allow access to company resources from any geographical location.
    \item The cloud provider taking over management of an organisation’s IT infrastructure allows for it to focus on its core business. 
\end{itemize}

\subsection{Cloud Delivery Models}
\subsubsection{Infrastructure-as-a-Service (IaaS)}
IaaS presents the cloud provider’s machines as a consumable item. Hardware is made available as virtual machines with facilities for processing, storage, and memory made configurable to preferences. Moreover, new  instances running on varying operating systems can be developed, enhancing the flexibility for distributed environments. Examples of such platforms are Amazon Elastic Cloud Compute (EC2) by Amazon Web Services and DigitalOcean. \cite{hajibaba, wadhe}

\subsubsection{Platform-as-a-Service (PaaS)}
PaaS allows consumers to focus on deployment and management of applications, lifting off functionalities such as resource management and planning, operating system maintenance, and security enforcement are lifted off and the consumer is only concerned with how to build his applications. \cite{devi} Examples of PaaS are Heroku and Azure App Services by Microsoft Azure, where the platforms come with development environments and other tools for programming such as Version Control, IntelliSense and debuggers. Services for Continuous Integration and Continuous Delivery (CI/CD) such as Bitbucket, Kubernetes and Travis CI can also be integrated to PaaS platforms to facilitate incremental code changes in the software development life-cycle. \cite{karamistos}

Besides application development, other services managed over PaaS are:
\begin{itemize}
    \item Database-as-a-Service (DBaaS) - for management of databases including licensing, upgrades and performance.
    \item Internet of Things (IoT) Platforms - such as Thingspeak, IBM Watson IoT and Blynk facilitate aggregation, analysis and visualisation of data from numerous interconnected devices. \cite{debauche}
    \item Mobile Services - some platforms allow organisations to transact with customers via their mobile devices, for instance, an automated reply system over SMS or IM. They can be used in mass communication and even individual inquiries, forming the Communications-Platform-as-a-Service (CPaaS). \cite{rcr}
    \item Machine learning (MLaaS) - platforms support a variety of frameworks such as TensorFlow, XGBoost and Pytorch to carry out different AI applications such speech and text processing, recommendation and ranking systems and computer vision. \cite{pawar} 
    \item Serverless computing - similar to application-oriented PaaSs, serverless computing platforms operate at a higher level of abstraction, in that developers are only concerned about the functionality of applications while details on infrastructure are handled by the cloud provider. Platforms such as Amazon Lambda, Google Cloud Functions and Microsoft Azure Functions are offered as a pay-as-you-go billing model and allow for automatic scaling. \cite{shafiei, sanders}  
\end{itemize}

\subsubsection{Software-as-a-Service (SaaS)}
SaaS are consumer-ready applications developed and hosted by a cloud provider, eliminating the need for on-premise installation, allowing for remote worker access to company resources. \cite{marston} These are popular end-user applications such as email services, productivity suites such as Google Workspace and Microsoft Office and enterprise-level services such as Salesforce, for customer relationship management.  

\begin{figure}
\centering
\includegraphics[width=\columnwidth]{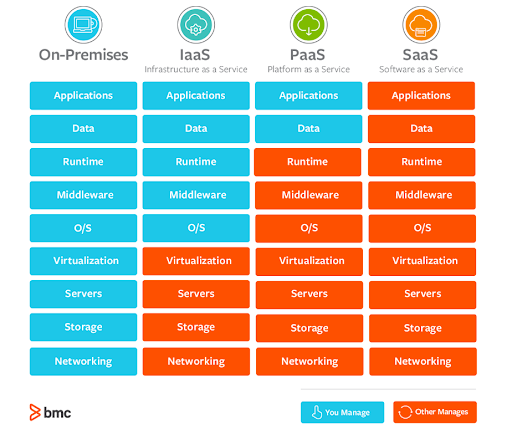}
\caption{Service Management in On-Premises and Cloud Delivery Models \cite{watts-raza}}
\label{fig:cd-model-services}
\end{figure}

Figure \ref{fig:cd-model-services} summarises how different system services are managed by cloud delivery models, compared to on-premises setups. In on-premise environments, services are implemented and controlled by system managers unlike SaaS, where management is done by cloud providers.

\subsection{Distributed Cloud Architecture}
\begin{figure}
\centering
\includegraphics[width=0.95\columnwidth]{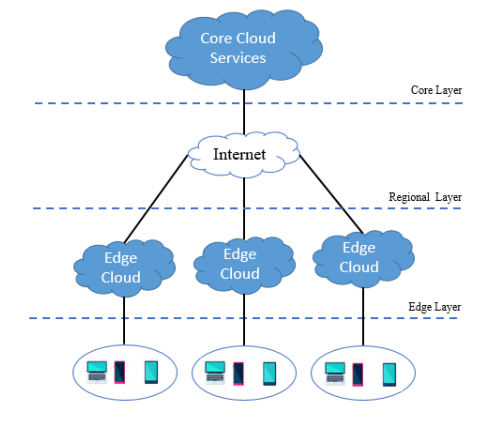}
\caption{Distributed Cloud Architecture \cite{moutai}}
\label{fig:dist-cloud-arch}
\end{figure}

Distributed cloud computing involves incorporating multiple cloud environments located in different geographical locations and managing them from a single public cloud provider. It is highly beneficial to organisations that aim to meet performance and usability requirements in their applications, comply with regulatory requirements, and avoid downtime in case of instances of data centre interruptions. \cite{felemban, pradhan, debauche} Design of these architectures is greatly influenced by how users, in a decentralised manner, discover and access underutilised resources. \cite{khethavath} Figure \ref{fig:dist-cloud-arch} represents a distributed cloud architecture, with three sections: core cloud, that includes all cloud management and provisioning services; regional cloud, that enhances communication and information sharing between layers; and edge cloud, which is in direct connection and communication with the customers. \cite{moutai}

Applications in Internet-of-Things, edge computing and over the Internet multimedia operate efficiently in distributed cloud environments as the load is conveniently distributed from the core to edge and regional clouds, guaranteeing high quality-of-service. \cite{nguyen, pradhan, debauche} Moreover, user mobility is enhanced, since the path of a cloud service can change among the edge clouds which eventually connect to the core cloud, achieving location transparency. \cite{nguyen} Some concerns regarding the architecture that can be addressed are meeting bandwidth requirements for the entire system that can affect quality of service and real-time performance, security measures for resources that are scattered across nodes and are easily accessible, as well as heterogeneity of infrastructures, platforms, and network technologies. \cite{felemban, khethavath, nguyen}

%% file: sections/cloud-db.tex
\section{Cloud Distributed Database Case Study: AWS Relational Data Service}
\label{cloud-db}
\subsection{Introduction}
As mentioned in the previous section, database services on the cloud are offered as DBaaS, that allow for easy setup and management of databases without needing to worry about aspects such as scalability, maintainability, and fault tolerance, which can affect the performance of a running system. \cite{munir} Besides transactional operations, cloud databases can also be tailored as big data solutions for business analytics and decision support systems. \cite{shah} Cloud providers have taken up an approach in designing distributed data stores, where concepts of database partitioning and replication are combined, enhancing their availability and performance. \cite{arias} Characteristics of such setups are \cite{kaur, kleppmann}:
\begin{itemize}
    \item Partitioning of data is done across multiple nodes with no partition having similar data with another. A partition key is applied to distribute the data.
    \item Each fragment has limitations in terms of bandwidth; implying that when partitions are created, it’s crucial to avoid one partition getting clustered by transactions while others remain under-utilised. To deal with issues arising from clustering, more partitions can be added to split the traffic.
    \item Data is replicated across multiple hosts or data centres, greatly increasing its availability. Different algorithms are implemented to determine how changes are applied across all replicas, considering issues such as network latency and tolerance of host faults.
\end{itemize}
Amazon Relational Data Service (RDS) is a collection of managed cloud database engines provided by Amazon Web Services (AWS), that allow for simple setup, operability, and scaling of distributed databases. The cloud engines include Amazon Aurora with MySQL, MYSQL, PostgreSQL, Oracle, and SQL Server as well as Amazon RDS with AWS that enable a hybrid deployment model. Numerous features make this service stand out \cite{aws-a, rapyder}: 
\begin{itemize}
    \item Lower administrative burden - tasks such as software versioning and licensing are easily handled by use of management consoles and RDS command line interfaces.
    \item High performance - the provisioned storage uses SSD devices that deliver a consistent baseline of 3 Input/Output Operations per second (IOPS) per provisioned GB and a burst of 3,000 IOPS under high loads.
    \item Scalability - Memory and compute resources are scalable with these upgrade procedures consuming a few minutes and minimal downtime. 
    \item Availability and durability - Automated backups are always enabled and data is retained for up to thirty-five days. Database snapshots further provide more resilience and are only deleted explicitly.
    \item User-friendly management - Database management is made efficient with monitoring and metrics, and event notifications. Metrics such as compute/memory/storage capacity utilisation, I/O activity, and instance connections allow database administrators to quickly detect performance problems. 
\end{itemize}

\subsection{Comparisons with Similar Cloud Distributed Database Systems}
AWS RDS has previously been compared to other similar services such as Azure Cosmos DB, Azure Database for MySQL, Azure Database for PostgreSQL by Microsoft Azure, and Google Cloud SQL, Cloud Firestore, and Google Cloud BigTable from Google Cloud. \cite{curino}

For this review, Amazon RDS is compared against Microsoft Azure SQL Database and CockroachDB by Cockroach Labs, which are all widely applied cloud based services for relational distributed databases, as in Table 2. \cite{crc-b, oracle, richard} The first two are cloud native databases whereas CockroachDB is offered as an open source service that is plugged in to a cloud service.

\begin{table*}[!t]
  \centering
  \caption{Characteristics of Featured Cloud Distributed Databases}
  \begin{tabular}{|p{2cm}|p{4cm}|p{4cm}|p{4cm}|}
        \hline
            \textbf{Feature}&\multicolumn{3}{|c|}{\textbf{Database Service}} \\
            \cline{2-4} 
            \textbf{} & \textbf{\textit{Amazon RDS}}& \textbf{\textit{Microsoft Azure SQL Database}}& \textbf{\textit{CockroachDB}} \\
        \hline
            Storage Engines & 
            Supports engines for Oracle, MariaDB, MySQL, Amazon Aurora, PostgreSQL & 
            Supports engines for SQL, PostgreSQL, MySQL such as InnoDB & Supports Pebble – a key-value store based on SQL and PostgreSQL  \\
        \hline
            Mode of Fragmentation & 
            Uses index-based partitioning for SQL based servers \cite{aws-c, vlasceanu} &
            Applies indexes and page density techniques \cite{msft-c} &
            Partitions established by grouping rows or ranges of rows \cite{crc-e} \\
        \hline
            Mode of Replication &
            Uses identifiers for MySQL and binary log file positions for MySQL and MariaDB instances to create up to 5 replicas \cite{aws-b} &
            Offers transactional replication to capture incremental changes as they occur and snapshot replication for infrequent changes to the database. \cite{msft-20, msft-21} &
            Carries out triplication storing each replica in its own node by default \cite{crc-d} \\
        \hline
            Cost Management &
            Billing is on the hardware resources used i.e., VCPUs, RAM, Storage, network bandwidth. &
            Billing is based on database size, concurrent connections, and throughput levels. &
            It is open source for application in a single region cloud (Serverless), multi-region cloud (Dedicated) and multiple clouds and regions (Self-Hosted) \\
        \hline
            Database Security &
            Security is guaranteed through service generated encryption certificate in the KMS (Key management service) &
            Azure Databases have security by the encryption certificate generator called Azure Key Vault service. &
            Provides role-based access control, client authentication and encryption based on the environment \cite{crc-c} \\
        \hline
            Disaster Recoverability and Availability &
            Highly available multi A-Z configurations against localised failures with availability of up to 99.95\% \cite{eisenberg} &
            The availability provided is up to 99.995\% with read replicas always available. \cite{eisenberg} &
            The model is a multi-active availability that keeps applications online in event of failure. \cite{crc-a} \\
        \hline
    \end{tabular}
\end{table*}

\subsection{Database Management with AWS RDS}

To demonstrate how databases are managed, a sample walkthrough is provided covering how a MySQL database is created, provisioned, accessed, and monitored by the AWS RDS service.

\subsubsection{Creating the Database}
This is achieved by the command in figure 10 which is broken down as follows: 

\begin{figure}
\centering
\includegraphics[width=\columnwidth]{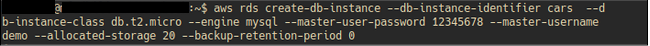}
\caption{CLI Creation of AWS RDS Database}
\label{fig:rds-create}
\end{figure}

\begin{itemize}
    \item \texttt{aws rds} : calls Amazon Web Services Relational Data Service
    \item \texttt{create-db-instance} : creates a new RDS database
    \item \texttt{--db-instance-identifier cars} : names created instance to ‘cars’
    \item \texttt{--db-instance-class db.t2.micro} : sets the computing and memory capacity to specified AWS capacity classes. The \texttt{t2.micro} class has 1 virtual CPU provided with 1 GB memory and limited networking capabilities. \cite{aws-b}
    \item \texttt{--engine mysql} : sets database engine to MySQL
    \item \texttt{--master-user-password 12345678} : creates password for master user to 12345678
    \item \texttt{--master-username demo} : creates and sets master user with username ‘demo’
    \item \texttt{--allocated-storage 20}: allocates 20GB database storage
    \item \texttt{--backup-retention-period 0} : number of days retained for automated backups, setting the parameter to 0 disables automated backups
\end{itemize}

\subsubsection{DBaaS Infrastructure Provisioning} 
\begin{figure}
\centering
\includegraphics[width=0.8\columnwidth]{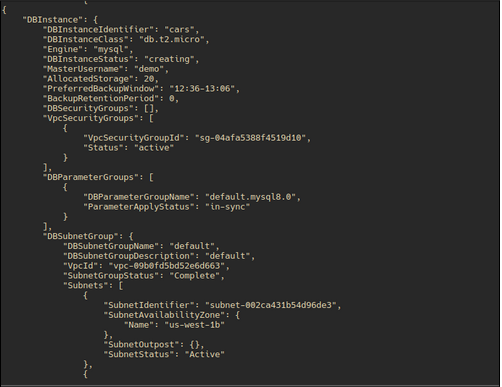}
\caption{AWS RDS Database Provisioning}
\label{fig:rds-provisioning}
\end{figure}

In the process of database creation, database infrastructure i.e., hardware resources are allocated by the PaaS service. Upon successful execution of the command in Figure \ref{fig:rds-create}, the result is as in Figure \ref{fig:rds-provisioning} which indicates resources such as subnets (\texttt{DBSubnetGroup}), virtual private cloud for security (\texttt{VpcSecurityGroups}) and storage (\texttt{AllocatedStorage}) are indicated. 

\subsubsection{Client Access and Manipulation}
\begin{figure}
\centering
\includegraphics[width=0.8\columnwidth]{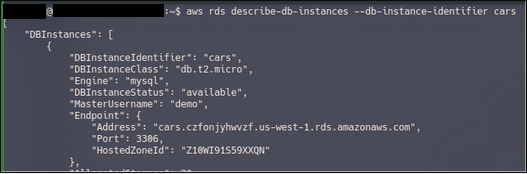}
\caption{Retrieving RDS Database Instance Details}
\label{fig:rds-retrieval}
\end{figure}

Once a database instance is up, it can be accessed by end clients for manipulation. First, application developers need to determine the database’s address. This is done by entering the command : \texttt{aws rds describe-db-instances --db-instance-identifier cars}, leading to Figure \ref{fig:rds-retrieval} as the result.

\begin{figure}
\centering
\includegraphics[width=\columnwidth]{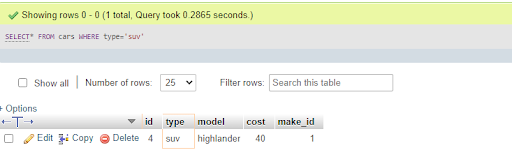}
\caption{Querying AWS RDS Database}
\label{fig:rds-query}
\end{figure}

The address is provided in \texttt{Endpoint-Address}, which is then included in end-user applications to direct queries. Thereafter, users can run standard SQL queries such as that in Figure \ref{fig:rds-query} that returns records from the cars database based on a value matching ‘suv’ in the ‘type’ attribute.

\subsubsection{AWS View Database Statistics}
\begin{figure}
\centering
\includegraphics[width=\columnwidth]{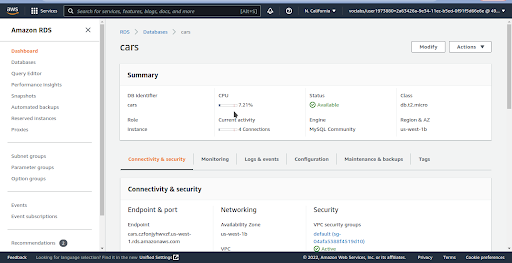}
\caption{AWS RDS Management Console for Database Statistics}
\label{fig:rds-mgt}
\end{figure}

On the AWS management console, as illustrated in Figure \ref{fig:rds-mgt}, that is accessible by the system administrator, they can determine different metrics that affect the performance of the database such as number of connections, physical location, and processor usage. The console also includes other database management functions such as backup and recovery, security logs and network configuration. 

%% file: sections/conclusion.tex
\section{Conclusion}
Distributed cloud database environments are critical in enhancing business operations. By partitioning data to be availed in several environments, they reduce the time for processing requests, enhancing the customer experience and overall system performance. They also carry out effortless management of database services and enable decision support systems through the setup and management of data warehousing facilities.

In describing the technology behind cloud distributed databases, this paper has illustrated the concepts that support its architecture i.e., distributed database systems and models for delivering services over the cloud. For the former, it has covered the design concerns and concepts that drive architecture design and implementation, together with the process of transaction management. Cloud based architectures and service delivery models were looked into, since they directly influence the application of cloud distributed databases. Eventually, a sample implementation of Amazon Relational Data Service, a renowned distributed database service was provided to showcase the creation, manipulation, and management of a relational distributed database. 

While implementing cloud-distributed database environments, it is crucial to understand issues around system security, scalability of data and resources and latency during transaction processing. Extensive research and development has been done to resolve these and many other concerns, therefore businesses also need to also include plans around the sustainability of cloud-distributed database environments within their deployments, to ensure optimal system performance.